\documentclass{emulateapj}

\def\vel{km s$^{-1}$}

\slugcomment{Re-submitted on 25 November 2008}

\shorttitle{\sc vibrationally excited CO}
\shortauthors{\sc Patel et al.}

\begin{document}


\title{Detection of Vibrationally Excited CO in IRC$+$10216}


\author{Nimesh A.\ Patel\altaffilmark{1}, Ken H.\ Young\altaffilmark{1},
Sandra Br\"unken\altaffilmark{1,2}\\  Karl
M.\ Menten\altaffilmark{3},
Patrick Thaddeus\altaffilmark{1},
Robert W.\ Wilson\altaffilmark{1}} 

\altaffiltext{1}{Harvard-Smithsonian Center for Astrophysics,  60 Garden
Street, Cambridge, MA; npatel@cfa.harvard.edu}
\altaffiltext{2}{Laboratoire de Chimie Physique Mol\'{e}culaire, Ecole Polytechnique F\'{e}d\'{e}rale de Lausanne, EPFL SB ISIC LCPM, Station 6, CH-1015 Lausanne, Switzerland}
\altaffiltext{3}{Max-Planck-Institut f{\" u}r Radio Astronomie, Auf dem H{\"
u}gel 69, D-53121, Bonn, Germany}

\begin{abstract}
Using the Submillimeter Array we have detected the $J = 3~-~2$ and
$2~-~1$  rotational transitions from within the first vibrationally
excited state of  CO toward the extreme carbon star IRC+10216 (CW
Leo).  The emission remains spatially unresolved with an angular
resolution of $\sim 2''$ and, given that the lines originate from
energy levels that are  $\sim3100$ K  above the ground state, almost
certainly originates from a much smaller ($\sim 10^{14}$ cm) sized
region close to the stellar photosphere.  
Thermal excitation of the
lines requires a gas density of $\sim10^{9}$ cm$^{-3}$, about an
order of magnitude higher than the expected gas density based
previous infrared observations and models of the inner dust shell
of IRC+10216.  \end{abstract}


\section{Introduction}
Emission from a rotational line from within a vibrationally excited
state of a molecule was first accidentally discovered by Snyder and
Buhl (1974), who discovered very strong maser emission from
vibrationally excited SiO toward the Orion Kleinmann-Low star-forming
region. Over the three past decades, SiO maser emission in the $J
= 1 - 0$ rotational line from the \textit{v} = 1 and 2  states (and
the \textit{v} = 1, $J = 2 - 1$ line) was found to be a ubiquitous
attribute of oxygen-rich mass-losing long period variable stars
(e.g., Messineo et al. 2002 and Deguchi 2007).

Following the first detections of vibrationally excited SiO maser
emission, it was immediately conjectured that similar emission---perhaps
even more intense---might be observable from the far more abundant,
spectroscopically similar CO molecule.  Fairly soon, Scoville and
Solomon (1978) detected a line in the carbon-rich, very high mass-los
AGB object IRC+10216 at exactly the frequency of the CO J = 1--0
transition in the first excited \textit{v} = 1 vibrational state.
This quite plausible assignment however fell through when it was
shown that the line in question was produced not by CO, but by the
C$_{4}$H radical, one of the more abundant molecules in IRC+10216
(Cummins et al. 1980; Gu\'elin et al. 1978).  In the intervening
30 years, no credible claim of vibrationally excited CO in the radio
spectrum of this evolved carbon star has appeared.  This apparent
difference between SiO and CO has yet to be explained.

The recent discovery with the Submillimeter Array\footnote{The
Submillimeter Array is a joint project between the Smithsonian
Astrophysical Observatory and the Academia Sinica Institute of
Astronomy and Astrophysics, and is funded by the Smithsonian
Institution and the Academia Sinica.} (SMA, Ho et al. 2004) in
IRC+10216 of vibrationally excited rotational lines from several
molecules (Patel et al. 2008) has reopened the question of vibrationally
excited radio lines and led --- we show here --- to the discovery
of vibrationally excited CO in two rotational transitions.  The new
data are summarized in Figures 1 and 2 and Table 1.  There can be
little doubt that the assignment is correct, because the probability
of a chance coincidence of unrelated lines is significantly less
than  $10^{-2}$.  A confirming transition, $J = 6 - 5$ at 685.18
GHz is accessible to the SMA, but requires low atmospheric opacity
to observe.

\section{Observations and Data reduction}\label{obs}

IRC$+$10216 was observed with the SMA as part of a spectral-line
survey in the 300--350 GHz ($0.87~\mu$m) band (Patel et al. 2008).
The CO \textit{v}=1, J=3--2 line was detected from observations
carried out on 2007 Feb 12, with a center frequency of 344.4 GHz
in the upper sideband. The array was in the so-called subcompact
configuration,  with baselines from 9.5 m to 69.1 m. The zenith
optical depth at the radiometer monitoring frequency of 225 GHz
($\tau_{225 {\rm GHz}}$) was about 0.05 and the double sideband
system temperature ($T_{\rm sys}$) varied from 130 to 260 K. Follow
up observations to confirm the identification by detecting the
J=2--1 vibrationally excited line, were done on 2008 June 5, with
the SMA  in the `extended' configuration, at baselines from 16.4
to 139.2 m. T$_{sys}$ varied from 70 to 250 K.  These observations
were done  during the day with $\tau_{225 {\rm GHz}}$ varying from
0.1 to 0.2. Pointing offsets were measured three times during the
observations by observing Mars and 3C273.  The receivers were tuned
to 228.4 GHz in the upper sideband. The phase center was   $\alpha,
\delta(J2000)=09^{\rm h}47^{\rm m}57\rlap{.}^{\rm s}38,
=+13^{\circ}16'43\rlap{.}''70$ for all the observations. Complex
gain calibration was done using the quasars 0851+202 and 1055+018;
the total observing time was about 8 hours.  The spectral band-pass
calibration was determined from observations of Mars and Jupiter
for the 2007 Feb 12 observations, and 3C279 for those on 2008 June
5. Absolute flux calibration was obtained  from observations of
Callisto.  The quasars  0851+202 and 1055+018 were observed every
20 minutes for complex gain calibration. The frequency resolution
was 0.81 MHz and 0.4 MHz for the 2007 Feb 12 and 2008 June 5
observations, respectively. The spectra shown here are smoothed to
an effective velocity resolution of 1 \vel.

The visibility data were calibrated using a standard package ({\it
Miriad}) for reduction of interferometric observations  (Sault,
Teuben \& Wright 1995).  The synthesized beam was $2.''9\times2.''2$
on 2007 Feb 12  and $3.''8 \times 1.''4$ on 2007 June 5.  Visibilities
were corrected for the position offset of
($\Delta\alpha,\Delta\delta)\approx(0.''7,0.''2)$ from the phase
center position, determined from the peak of continuum emission
(Patel et al. 2008).  All the spectra here were produced by spatially
integrating the continuum-subtracted line intensity in a $2''\times2''$
rectangle centered on the peak continuum emission.

\section{Results}\label{results}

Figure 1 shows the CO  \textit{v} = 1, $J = 3 - 2$ and $J = 2 - 1$
spectra toward IRC+10216.  The dashed line indicates the systemic
velocity of $-26.2$ \vel (Olofsson et al. 1982).  Both lines are
slightly blue-shifted by 1--2 \vel relative to the star.  The spectra
were produced by integrating the continuum-subtracted line intensity
in a $2''\times2''$ rectangle centered on the continuum peak.  An
unrelated line at $\sim$11 \vel which appears in the the top panel
of Fig. 1 is probably CC$^{34}$S 8(7)--5(6) at  a rest frequency
of 342629.24 MHz.  That identification for this line is a plausible
choice, since (much stronger) emission in multiple lines of the
main isotopologue CCS have been identified in IRC+10216 previously
(Cernicharo et al. 1987).  The frequency of the  CO \textit{v} =
1, $J = 1 - 0$ transition is very close to that of the $N,J = $
$12,25/2 - 11,23/2$ line of the C$_{4}$H radical (Cummins et al.
1980), making it difficult to pick out the vibrationally excited
CO line in the 3 mm band with a single-dish telescope. The CO
\textit{v} = 1 $J = 3 - 2$ line on the other hand, appears free
from this confusion, although it may be blended with CH$_{3}$C$_{3}$N
less than a MHz away (at $-0.7$ km s$^{-1}$ in the spectrum shown
in Fig. 1) ---  a molecule however, not yet detected in IRC+10216
(Agundez et al. 2008).

Crucial confirmation of the present identification is provided by
detection of the vibrationally excited J=2--1 line, which is free
from confusion with other lines over a frequency interval of $\sim$
2 MHz.

\begin{figure}
\begin{center}
\includegraphics[width=3.5in]{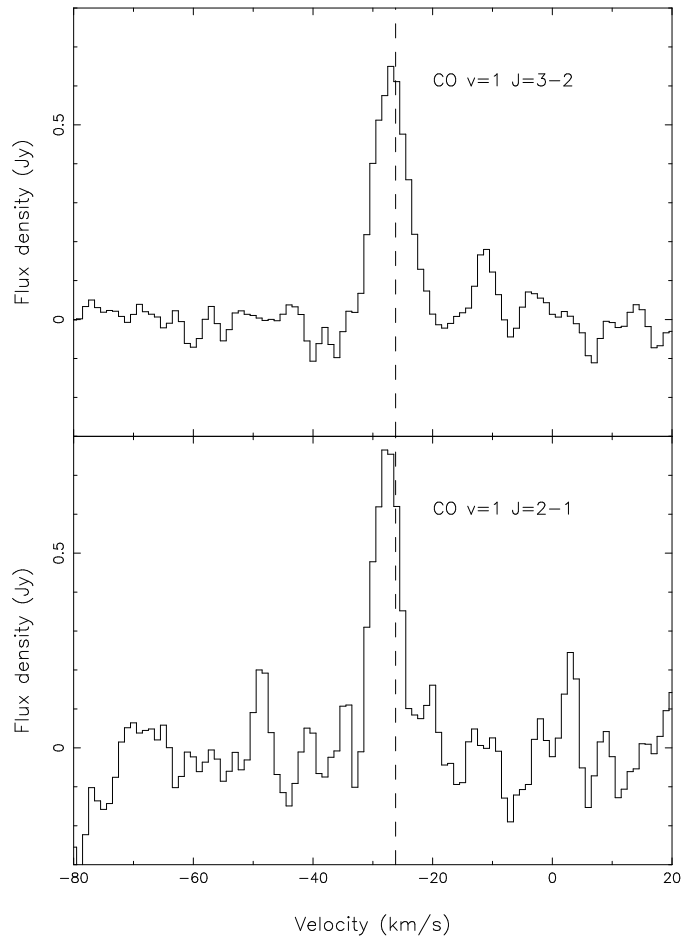}
\end{center}
\caption{Spectra of vibrationally excited CO emission in $J = 3 - 2$ (top) and
$J = 2 - 1$ (bottom) toward IRC+10216.} 
\end{figure}

Maps of the integrated intensity of the CO \textit{v} = 1, $J = 3
- 2$ and $J = 2 - 1$ emission are shown in Figure 2. The position
offsets are with respect to the peak of the continuum emission
quoted above.  The centroid of the 3--2 and 2--1 line emission
distribution is coincident with that of the continuum emission.
Gaussian fitting and deconvolution with the {\it IMFIT} task of the
Miriad package yields a size of $1.''6\times0.''9$ with a P.A of
$-28^{\circ}$ for the $3 - 2$ emission and a point source for the
$2 - 1$ emission. Visibilities plotted vs. $uv$ distance for both
lines do not show a tapering for large distance values, suggesting
that both the lines are spatially unresolved. The apparent
``resolution'' of the $3 - 2$ emission distribution is thus almost
certainly caused by broadening due to residual phase calibration
errors.

\begin{figure}
\begin{center}
\includegraphics[width=3.5in]{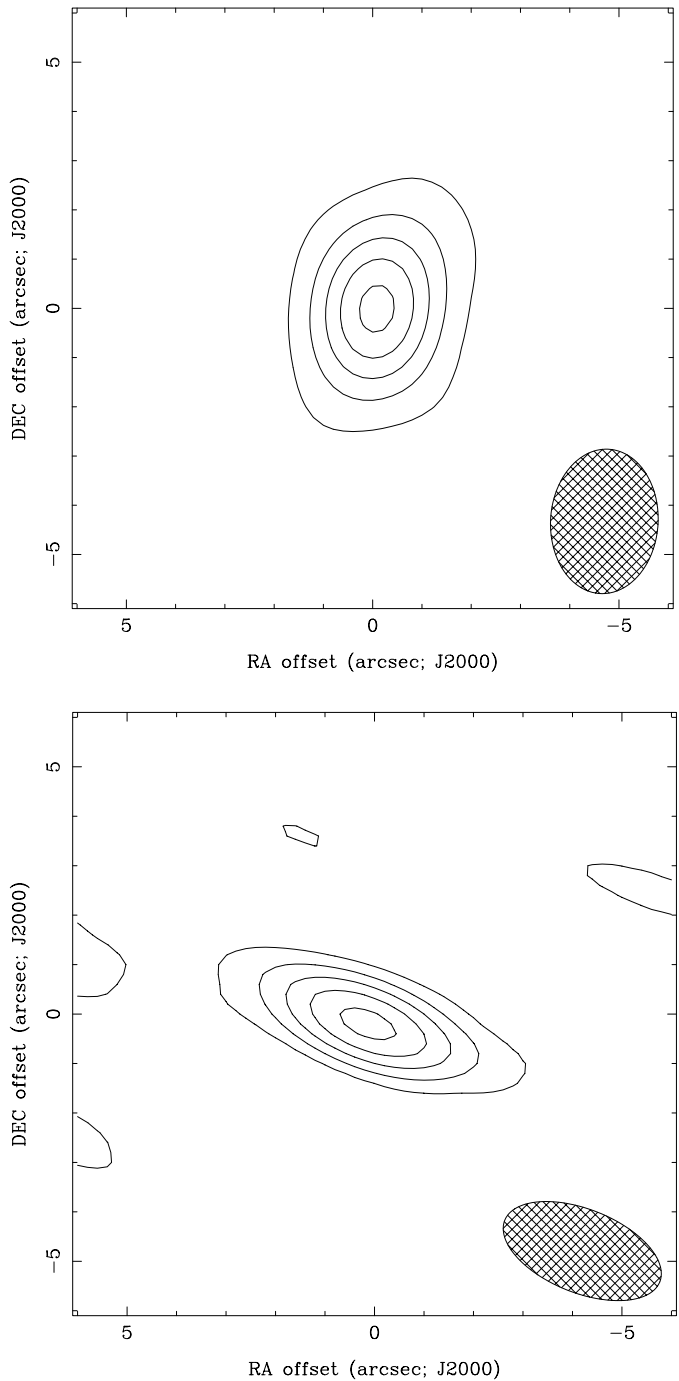}
\end{center}
\caption{Integrated intensity maps of CO \textit{v} = 1 $J = 3 -
2$ (top) and $J = 2 - 1$ (bottom) emission. Contour level intervals
and starting value are 5$\sigma$ with $\sigma=0.16$ Jy beam$^{-1}$
\vel in both panels. The synthesized beam is shown in the lower
right corner. The emission is spatially unresolved. } \end{figure}

\section{Discussion}\label{discussion}

Consistent with the fitting results, maps of the \textit{v} = 1 $J
= 3 - 2$ and $J = 2 - 1$ CO emission (Figure 2) show very compact
spatial distribution.  The vibrationally excited CO emission is
expected to occur very close to the star since the relevant energy
levels are 3100 K above the ground state.  The vibrationally excited
CO lines have narrow line-widths implying V$_{exp}\la4$ \vel. Several
such narrow lines have been detected from our on-going spectral-line
survey of IRC+10216 (Patel et al. 2008) and some of these lines
have been re-observed with the SMA at higher angular resolution of
$\sim 1''$. For example, CS v=2 J=7--6 emission is spatially
unresolved with a deconvolved source size $\le 0.''2$.  We assume
in the present analysis that both the lines arise from the same
region of radius $0.''1$, which corresponds to $2.2\times 10^{14}$
cm (for a distance of 150 pc). We note that this is comparable to
the size of the inner radius of the dust shell in IRC+10216 deduced
from interferometric IR observations by Danchi et al. (1994).

For a Gaussian source size of $0.''2$ FWHM, from the observed flux
density of the 342.6 and 228.4 GHz CO lines, we obtain a brightness
temperature of 210 and 412  K, respectively.  Assuming Local
Thermodynamic Equilibrium (LTE), this would imply  optical depth
of 0.24 and 0.53, respectively, for the two lines. Danchi et al.
(1994) measure the kinetic temperature of the dust shell to be 1356
K at this radius. Previous work by Sahai and Wannier (1985) uses a
value of 600 K for the inner dust shell's temperature (see also
Toombs et al. 1972). We assume the temperature of the dust to be
the same as that of  gas, which we take to be 1000 K, also suggested
by an extrapolation of the temperature vs.  radius plot derived by
Crosas and Menten (1997) from the analysis of multiple $J$ transitions
of CO in the outer parts of the shell.  With this temperature and
with the further assumption of uniform optical depth in the line,
using the Gaussian fitted line-widths and Einstein A coefficients
for the $J = 3 - 2$ and $J = 2 - 1$ transitions in \textit{v} = 1
state (from SLAIM), we derive the column densities in the \textit{v}
= 1 J=2 (J=1) levels to be $1.3\times 10^{17}$ ($3.7\times 10^{17}$)
cm$^{-2}$, where we have adapted the standard relation for column
density,  excitation temperature and optical depth for CO (see Eqn.
14.43 of Rohlfs and Wilson 2000).  The total CO column density
$N$(CO) is obtained from above using the full partition function
including vibrational and rotational levels (Townes and Schalow
1975),  and an abundance value of  [CO]/[H$_{2}$]=$3\times10^{-4}$
(e.g., Sahai \& Wannier 1985), we derive the H$_{2}$ column density
to be $\sim10^{24}$ cm$^{-2}$.

Taking the diameter of the emission region, $l$, as
$2\times2.2\times10^{14}$ cm we derive the gas density, $n$(H$_2 =
N$(H$_2/l$), of $5\times10^{9}$ cm$^{-3}$.  This is about an order
of magnitude greater than the gas density obtained by Danchi et al.
(1994) for IRC+10216's inner dust shell.

These rough estimates suggest that collisional excitation under the
LTE assumption is unlikely to explain our observations. To check
whether the \textit{v} = 1 levels may be excited via radiation, we
use the condition: $(e^{{h\nu_{IR}\over kT}}-1)^{-1}>A_{rot}/A_{vib}$
(Carrol and Goldsmith 1981; Menten et al.  2006a). We have
$A_{rot}=2.4\times10^{-6}$ s$^{-1}$ (for the $J = 3 - 2$ transition),
$A_{vib}=11.5$ s$^{-1}$,  $\nu_{IR}=6.5\times10^{13}$ Hz  for 4.6
$\mu$m and T=1000 K, we find the ratio of the A coefficients to be
$2.2\times10^{-8}$ while the l. h. s. of the above expression has
a value of 0.27,  hence the v=1 levels are most likely  radiatively
excited. However, to attain such a high radiation temperature, the
emission must arise from a region even smaller than $0.''2$.

Since the observations of the $3 - 2$ and $2 - 1$ lines were carried
out at different phases of pulsation of IRC+10216 the relative
intensities of these two lines may complicate their interpretation.
We calculated the phase, $\phi$, for the two epochs of our observations,
$T_{\rm obs}$, given by $(T_{\rm obs}-T_{\rm max})/P)$, using 664
d as the period, $P$. For this we have calculated the time of the
last maximum considering the IR minimum date of 1989 Dec 5 given
by Danchi et al. (1994). We find $\phi=0.25 (0.99)$ for the 346
(228) GHz observations. These values of $\phi$ are in good agreement
with values obtained using the date of maximum IR flux (at 3.76
$\mu$m; Le Bertre 1992) on 1992 May 30 from cm- and IR-monitoring
reported by Menten et al. (2006b). Thus it is plausible that the
$J = 2 - 1$ emission is brighter due to its being observed closer
to the IR maximum, compared to our $J = 3 - 2$ observations.  Future
observations at different epochs may help checking the radiative
pumping hypothesis while near-simultaneous observations of different
$J$ transitions in \textit{v} = 1 state will be helpful for detailed
radiative transfer modeling of these lines.  While our present
observations have inadequate angular resolution to definitively
rule out maser action to explain the vibrationally excited CO lines,
we note that even for a $0.''1$ Gaussian source, the brightness
temperature would be 1451 (738) K for the 228.4 (342.6) GHz emission,
consistent with thermal emission. This would imply a gas density
$>10^{9}$ cm$^{-3}$.  We note that a strict lower limit to the size
of the CO emission is given by recent observations by Menten et al.
(2008) who derive a (uniform disk) diameter of  84 milli arcseconds
(mas) for IRC+10216 radio photosphere from their Very Large Array
measurements of the object's 7 mm wavelength continuum emission.

Scoville and Solomon (1978) proposed that the rotational lines in
\textit{v} = 2 state are pumped by stellar IR photons at 2.3 $\mu$m,
followed by rapid spontaneous decay to the \textit{v} = 1 state.
Analogous  to the Kwan and Scoville (1974) mechanism for the SiO
maser emission in \textit{v} = 1 state, the low rotational state
populations of \textit{v} = 1 state become inverted in the circumstellar
region where the \textit{v} = 1$\rightarrow$2 optical depth is low
and v=0 $\rightarrow$1 optical depth is high. There are differences
for CO, however, owing to its high abundance with respect to SiO
and low dipole moment. Deguchi and Iguchi (1976) proposed another
mechanism for SiO maser pumping that may also apply to CO masers.
According to their model, radiatively pumped \textit{v} = 1 levels
alone will invert the rotational populations if there is acceleration
in the region producing large velocity gradients. Both these
mechanisms require $J$-independent trapping of the \textit{v}$\rightarrow$
\textit{v}$-1$ transitions and require the IR optical depth to be
of the order of unity (Turner 1987).  The slightly blue-shifted
emission seen in both the lines is reminiscent of  the observations
by Turner (1987) of vibrationally excited SiS lines. He interpreted
the latter as examples of weak masers amplifying the stellar radiation
from the gas in the innermost expanded shell. While  a systematic
outflow will not have commenced inside the dust formation zone, the
blueshifts might nevertheless reflect outward motions of gas levitated
away from the star.

CO is so widespread in Galactic molecular gas, it would be surprising
if more sources of vibrationally excited emission could not be
detected.  Oxygen containing molecules are not especially abundant
in IRC+10216, except for CO, which ties up almost all of the oxygen.
Therefore, oxygen-rich sources like the red supergiant VY CMa may
turn out to be the best places to look. Toward this source very
many vibrationally exited lines from SiO have been found from up
to the 7000 K high v=4 state (Cernicharo et al. 1993) and from
H$_2$O from the $\nu_2=1$ bending mode (Menten \&\ Young 1995,
Menten et al. 2006a). Interesting in the context of our discussion
is that the latter authors found one of the H$_2$O lines they
detected consistent with thermal excitation, while weak maser
emission is suggested for the other.  Interferometric studies of
such objects are now clearly desirable. We have begun such a study
with the SMA and have observed R Cas and Mira. The CO \textit{v}=1
J=2--1 was not detected in either of these sources with an rms
noise level of 0.1 Jy/beam.

\section{Conclusions and Outlook}\label{conclusions}

We have reported the first unambiguous detection of (sub)millimeter
wavelength vibrationally excited CO lines in the $J = 3 - 2$ and
$2 - 1$ rotational transitions from \textit{v} = 1 state toward
IRC+10216.  Thermal excitation of these levels requires extremely
high gas densities and temperature, but the lines could be radiatively
excited. Population inversion in the levels cannot be ruled out,
but there is little evidence for it.

Observations of IRC+10216 with the eSMA\footnote{SMA combined with
the James Clerk Maxwell Telescope and the Caltech Submillimeter
Observatory} in the near future will provide an improvement in
angular resolution by nearly a factor of 10 over that here and will
probe vibrationally excited emission with $\sim 0.''2$ resolution.
In the future, the Atacama Large Millimeter Array (ALMA) will have
much higher angular resolution, up to 15 mas at 1 mm wavelength in
its longest (15 km) baseline configuration. Enabled by the excellent
brightness sensitivity afforded by it large collecting area, ALMA
will be able to \textit{image} the hot gas traced by vibrationally
excited lines. It will thus allow detailed studies of the complex
dynamics of the gas close to the stellar surface in which the
circumstellar outflow has hardly started, and on whose molecular
content now only line-of-sight averaged information from IR absorption
spectroscopy can be obtained (see Sahai \&\ Wannier 1985).

\acknowledgments

\small
\begin{deluxetable}{ccccccc}
\tablecolumns{7}
\tablewidth{0pt}
\tablecaption{Vibrationally excited CO line parameters\label{table1}}
\tablehead{
Transition&Rest
frequency\tablenotemark{1}&E$_{u}$/k&Peak\tablenotemark{2}&Integrated&Velocity\tablenotemark{3}
&V$_{exp}$\tablenotemark{4}\\
               &    (MHz)                                      &(K) & flux density  & flux
density  & (km s$^{-1}$)       & (km s$^{-1}$)         \\
                 &                                                  &      &  (Jy)            &
(Jy km s$^{-1}$) &                        &         }
\startdata
\textit{v} = 1 $J = 3 - 2$ & 342647.66$\pm$0.01 & 3116.6 & 0.71$\pm$0.05
&3.76$\pm$0.19&-27.03$\pm$0.26&3.79$\pm$0.06 \\
\textit{v} = 1 $J = 2-1$ & 228439.11$\pm$0.07 & 3100.1 & 0.62$\pm$0.04
&2.23$\pm$0.23&-27.86$\pm$0.21&3.58$\pm$0.67 \\
\enddata
\tablenotetext{1}{From the Spectral Line Atlas of Interstellar Molecules (c.f., Remijan et al. 2008).}
\tablenotetext{2}{These are 1$\sigma$ formal uncertainties from Gaussian fits. The actual
uncertainty in absolute flux calibration is $\sim$15\%.}
\tablenotetext{3}{Systemic velocity of the source $=-26.2$ \vel.}
\tablenotetext{4}{Half-width at half-maximum from Gaussian fits.}

\end{deluxetable}
\normalsize

\end{document}